\documentclass[conference]{IEEEtran}
\usepackage{cite}

\ifCLASSINFOpdf
\usepackage[pdftex]{graphicx}
\else
\fi
\usepackage{amsmath}
\usepackage{amsfonts} 
\usepackage{amsthm}
\usepackage{amssymb}
\usepackage{amsthm} 
\usepackage{mathtools} 
\usepackage{textcomp} 
\usepackage{xcolor}
\usepackage{clrscode}

\newtheorem{algorithm}{Algorithm}
\newtheorem{CodeListing}{Code Listing}

\usepackage{BOONDOX-ds}

\usepackage{bbm}

\newcommand{\pfun}{\mathop{\hbox{$\to$\kern-7pt\raise.9pt\hbox{\scalebox{1}[.55]{$|$}}\kern4pt} }}

\usepackage{array}

\begin{document}

\title{Easy Multi-Precision Acceleration \\ via Vectorization
}

\author{\IEEEauthorblockN{
Chansup Byun, Piotr Luszczek, Jeremy Kepner, \\ LaToya Anderson, William Arcand, David Bestor, William Bergeron, \\ Alex Bonn, Daniel Burrill, Vijay Gadepally, Michael Houle, Matthew Hubbell,  \\ Hayden Jananthan, Michael Jones, Lauren Milechin, Julie Mullen, Andrew Prout,  \\ Albert Reuther, Antonio Rosa, Charles Yee, Peter Michaleas
\\
\IEEEauthorblockA{
MIT
}}}
\maketitle
\begin{abstract}
High level programming languages and accelerators are powerful enablers for a wide range of applications.  Accelerators offer significant performance via diverse numeric precision and increased memory bandwidth and capacity.  These enable new algorithms to be explored.  Construction of these algorithms can be enabled by the ready accessibility of these performance features in high-level languages.  Achieving scalable vertical (within a compute node), precision (over various numeric precisions), and temporal (over different generations of hardware) performance while retaining productivity requires effective abstractions. Vectorized coding that expresses operations as high-level matrix mathematics has historically been an important method for achieving portable performance. This paper examines the performance availability across a wide range of precisions and hardware using simple vectorized coding techniques applied to several example programs that highlight  bandwidth, real and complex matrix multiplication, and 2D convolution. The systems evaluated span nearly a decade of computing hardware, enabling direct comparison of performance gains over this period.  This work demonstrates that near hardware peak performance is readily accessible from high-level programming environments.  In particular for memory bandwidth, real and complex matrix multiplication in double/single/half precision, and 2D convolutions using the Fast Fourier Transform (FFT).  Furthermore, algorithms that trade decreased compute intensity for increased memory intensity have significant potential. 

\end{abstract}


%
\IEEEpeerreviewmaketitle

\section{Introduction}
\let\thefootnote\relax\footnotetext{Research was sponsored by the Department of the Air Force Artificial Intelligence Accelerator and was accomplished under Cooperative Agreement Number FA8750-19-2-1000. The views and conclusions contained in this document are those of the authors and should not be interpreted as representing the official policies, either expressed or implied, of the Department of the Air Force or the U.S. Government. The U.S. Government is authorized to reproduce and distribute reprints for Government purposes notwithstanding any copyright notation herein.
}

\begin{figure}[bth]
\centerline{\includegraphics[width=1.0\columnwidth]{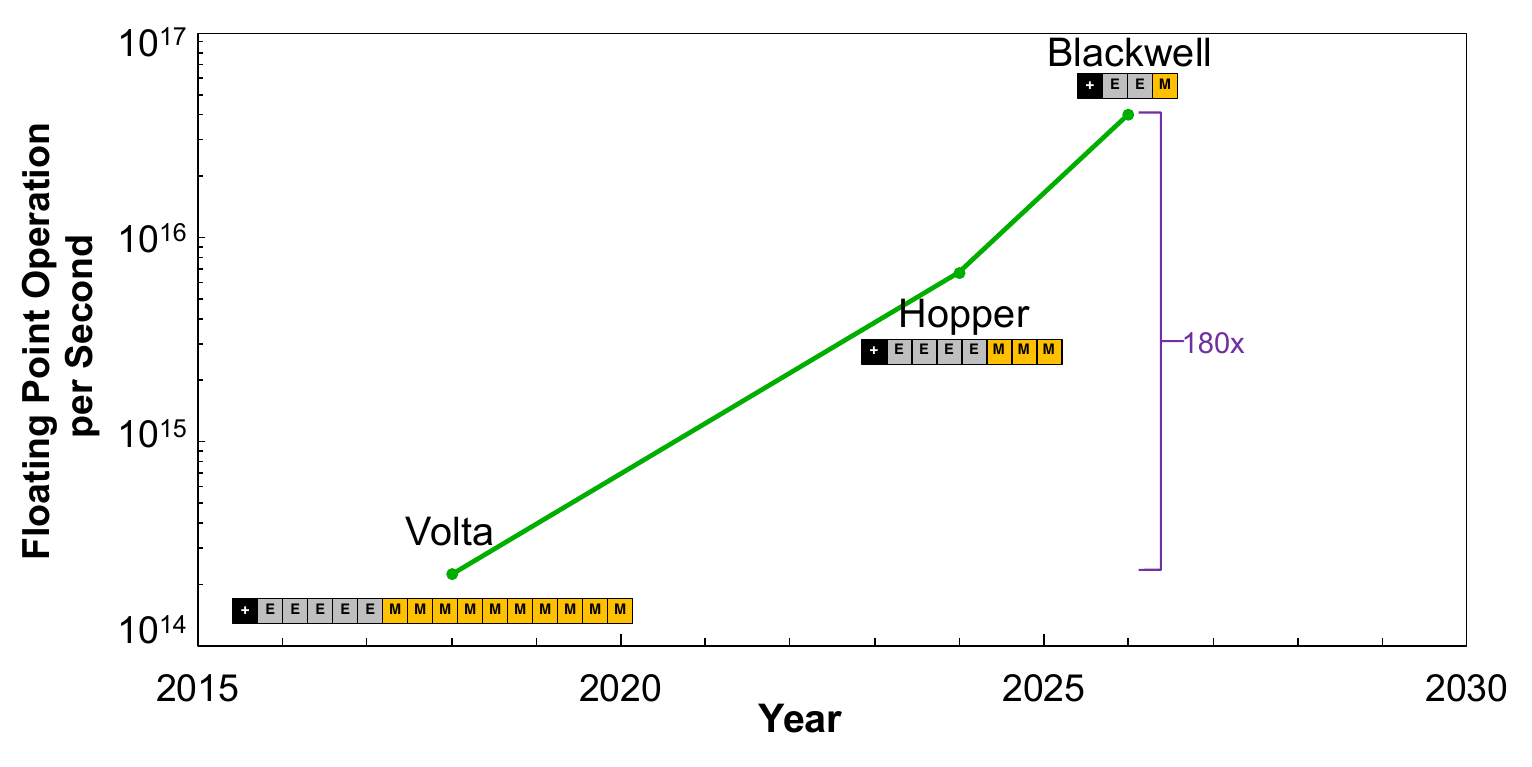}}
\centerline{\includegraphics[width=1.0\columnwidth]{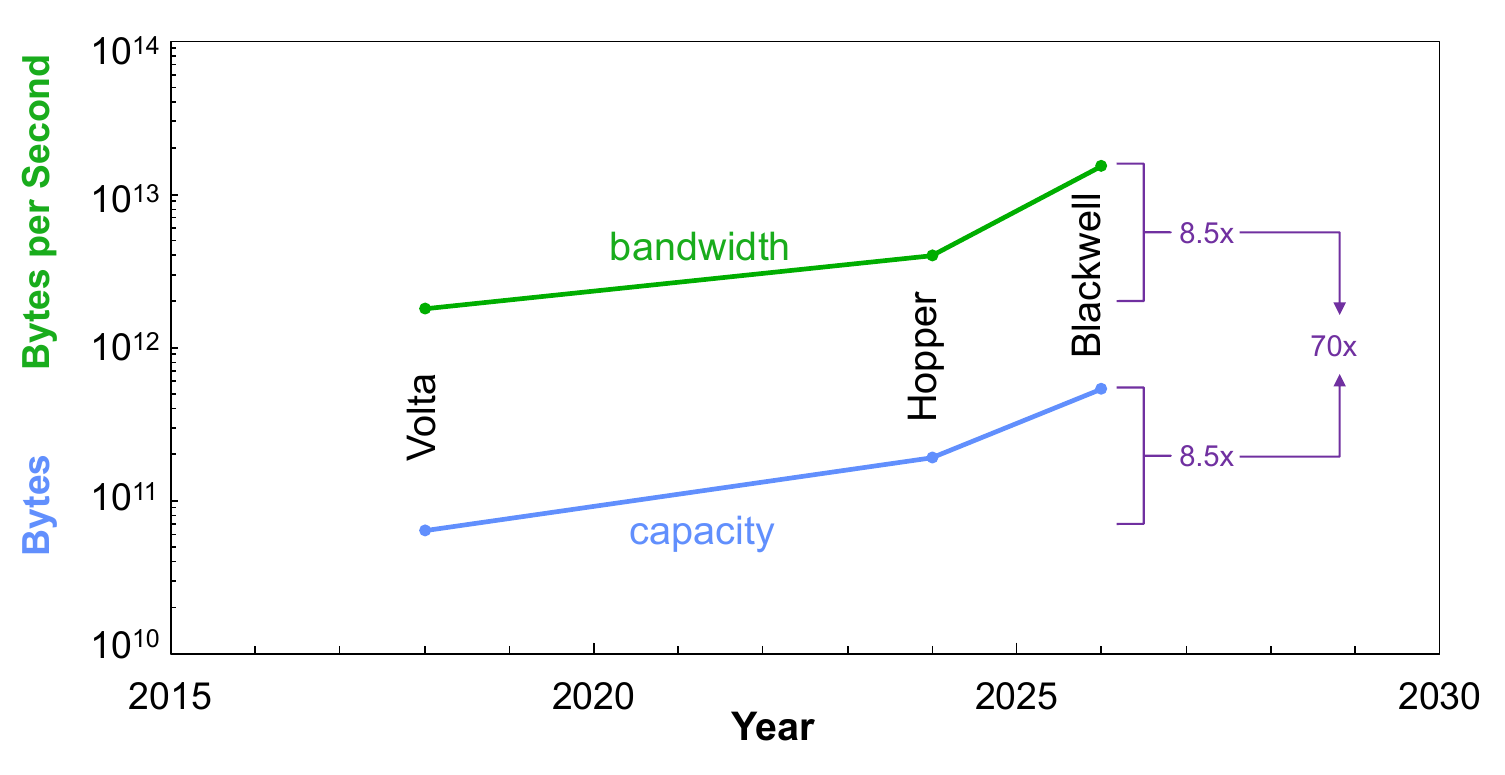}}
\caption{{\bf Precision, Performance, Bandwidth, and Capacity Trends.} Increasing performance of Nvidia GPUs as they offer optimized hardware for lower precisions (top) and larger memory (bottom). (top) + denotes a sign bit, E denotes an exponent bit, and M denotes a mantissa bit.
}
\label{fig:GPUTrend}
\end{figure}

Easy-to-use high level programming languages (e.g., Matlab \cite{higham2016matlab}, Python \cite{van1995python}, and Julia \cite{bezanson2017julia},) enable millions of programmers and play an important role in computational science and high performance computing.  Accelerators, most notably GPUs (Graphics Processing Units), have become key enablers for a wide range of computationally intensive workloads, particularly in AI (Artificial Intelligence) \cite{reuther2022ai, reuther2023lincoln}.  The capability of CPUs (Central Processing Units) continues to increase as the number of processing cores on a CPU grows.  Accelerators offer significant performance via diverse numeric precision and increase memory bandwidth. A key challenge is maintaining the ease-of-use of high level programming languages while achieving scalable performance across different dimensions: vertical (within a compute node), precision (FP64, FP32, FP16, ...), and temporal (over different generations of hardware).

Vertical scaling within nodes often relies on threading approaches (e.g., OpenMP \cite{dagum1998openmp}, pthreads \cite{nichols1996pthreads}, and Cilk \cite{blumofe1995cilk}) that leverage shared memory model.  Threading allows  libraries (e.g., MKL \cite{wang2014intel}, MAGMA\cite{agullo2009numerical}, SPIRAL\cite{franchetti2018spiral}, and GraphBLAS\cite{kepner2016mathematical, davis2019algorithm}) written in low-level languages (e.g., C\cite{kernighan1988c}, C++\cite{stroustrup2013c++}, and Fortran\cite{press1992numerical}) to deliver parallel performance at the function level that can be readily called by a high level programming language.  Vertical scaling on accelerators often follows a similar approach leverage technologies that are designed for more heterogenous accelerator architectures  (e.g., CUDA\cite{sanders2010cuda} and OpenCL\cite{munshi2011opencl}).

Accelerators have created a veritable zoo of precisions as they pursue the relatively low precision requirements of AI (see Figure~\ref{fig:GPUTrend}).  Added to the traditional double (FP64), single (FP64), and integer (int64, int32, int16, int8) types are half (FP16), quarter (FP8), eighth (FP4), brain float (BF16), and tensor float (TF32), which takes BF16 input and produces FP32 output).  The potential performance gains of these new types are significant and it is important for algorithm developers to have performant access to them in high level languages. 

Accelerators have also increased both memory bandwidth and memory capacity of their fastest programmable memories  (see Figure~\ref{fig:GPUTrend}).  The combined memory bandwidth and capacity improvements increasingly enables  algorithms that trade decreased compute for increased memory usage.  Specifically, algorithms that do fewer operations per byte but access those bytes more (e.g., sparse and spectral methods).

Vectorization \cite{cai2005performance, birkbeck2007dimension} is an important optimization technique that is particularly relevant to high-level languages that often use large arrays as a core construct.  Operating on large arrays can often be the key to leveraging the underlying high performance libraries that are built with the aforementioned vertical scaling methodologies.  Using vectorization to access accelerator performance begins with good memory bandwidth performance as high level languages can often use more memory than lower level languages.  The  Stream benchmark is a common benchmark for measuring bandwidth that vectorizes well \cite{McCalpin2005}.  This work builds on our prior work with this benchmark \cite{kepner2025easy}.

Diverse precision performance first requires the availability of the corresponding types in the high level language. Next, the extent that high level languages makes it easy to access the underlying hardware performance can be assessed.  Matrix multiply is a good tool for exploring both as matrix multiply is a defined mathematical operation for diverse types (both real and complex) and the underlying hardware is often design to accelerate this operation.  Together, Stream vector operations and matrix multiplication are a good foundation for building many real applications, particularly in the area of sensor processing. Two important sensor processing applications are beamforming and image convolution.

Beamforming is the basis of many sensing and communication applications, including: radar, mineral exploration, medical imaging (MRI, ultrasound, …), wireless communications, land/sea/undersea/air/space manned/unmanned navigation and traffic management, satellite imaging, weather monitoring, oceanography, and radio astronomy.  Correspondingly, we have a developed and published a ``Beamformer'' mini-application that exercises the most computationally intensive aspects of beamforming \cite{Kepner2009}.  Beamforming relies on complex matrix multiplication over diverse precisions and should track the corresponding underlying matrix multiplication performance.

Image (2D) convolution is a foundation of most image processing that is widely use in many of the same applications as beamforming.  2D convolution is also critical to convolutional neural networks (CNNs) widely used in AI applications.   Likewise, we have a developed and published a ``Blurimage'' mini-application that exercises the most computationally intensive aspects of 2D image convolution \cite{Kepner2009}.  Blurimage relies on 2D convolution over diverse precisions (both real and complex) and can be implemented with compute intensive direct methods and memory intensive spectral methods using the Fast Fourier Transform (FFT) \cite{cochran1967what, nussbaumer1981fast, liu2018efficient}.

This paper presents a brief case study on the performance of accelerated vectorized codes over diverse precisions in high level languages, using Stream, matrix multiplication, Beamformer, and Blurimage.   These benchmarks are implemented in Matlab and Python and leverage the pMatlab\cite{Kepner2009} and pPython\cite{byun2023ppython} distributed array libraries for multi-process parallelism when required.  Performance is measured on a invidual nodes of varying CPU and GPU hardware spanning almost a decade of technology.  The remainder of the paper is organized as follows. We first describe basic vectorization and parallelization. We then present an in-depth look at the matrix multiplication benchmark. Subsequent sections describe the application benchmarks, the benchmarking hardware, and the performance results, followed by conclusions and a discussion of potential future work.

\section{Vectorization \& Parallelization}

  The essence of vectorization in high level languages is removing ``for loops'' and replacing them with higher level operations on whole vectors or matrices that can take advantage of underlying hardware optimized libraries.  In the case of Stream a good example is replacing
\begin{codebox}
\zi     \For $i=1:N$
\zi       \Do
\zi           {\bf a}(i) = {\bf b}(i) + q*{\bf c}(i)
          \End
\end{codebox}
with
\begin{codebox}
\zi           {\bf a} = {\bf b} + q*{\bf c}
\end{codebox}
where {\bf a}, {\bf b}, {\bf c} are N element vectors and $q$ is a scalar.  In the case of the Stream benchmark, there is an also outer loop that is run $N_t$ times.  Similarly, matrix multiplication of $N{\times}N$ matrices {\bf A}, {\bf B}, {\bf C} can be vectorized by replacing
\begin{codebox}
\zi     \For $i=1:N$, $j=1:N$, $k=1:N$
\zi       \Do
\zi           {\bf C}(i,j) += {\bf A}(i,k) * {\bf B}(k,j)
          \End
\end{codebox}
with
\begin{codebox}
\zi           {\bf C} = {\bf A} * {\bf B}
\end{codebox}

  Code vectorization can dramatically increase performance, reduces code size, simplifies debugging, and is often highly portable across generations of hardware.  Furthermore, these operations work well across a wide range of types and can provide a simple way to access underlying accelerated hardware of performance.   For example, simply changing the type of the vectors or matrices is typically all that is required to leverage different underlying computing hardware. 

  When comparing CPU and GPU programming it is important to use optimal approaches for each.  A common difference is that CPUs often achieve maximum performance when many instances of the same program are run on a multi-core CPU, while GPUs often to achieve maximum performance with running one instance of a program per GPU.  To easily explore these differences requires a method of to easily run multiple instances of the programs.
  
\begin{figure}[t!]
\centerline{\includegraphics[width=\columnwidth]{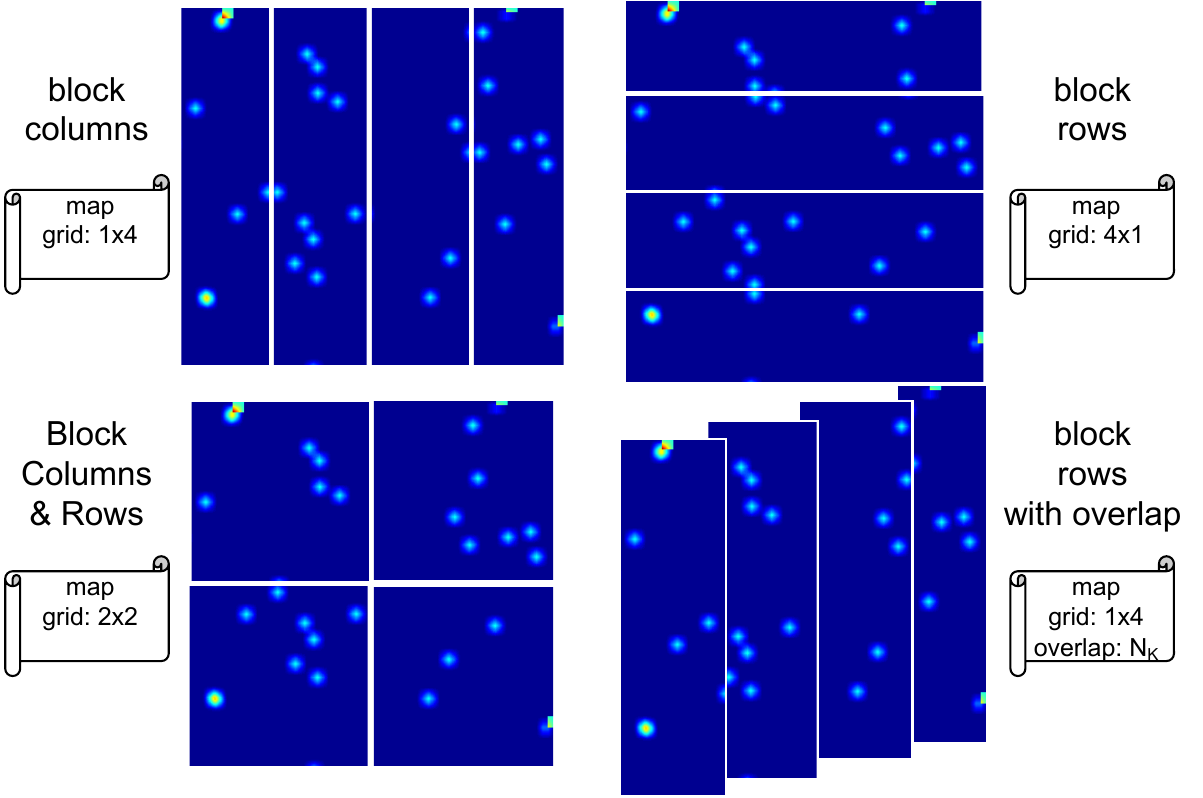}}
\caption{{\bf Distributed Array Mappings} (adapted from \cite{Kepner2009}).
Different parallel mappings of a two-dimensional array.  Arrays can be broken up in any dimension.  A block mapping means that each $P_{ID}$ holds a contiguous piece of the array.  Overlap allows the boundaries of an array
to be stored on two neighboring $P_{ID}$s.}
\label{fig:GlobalArrayMaps}
\end{figure}

  The distributed array model in many ways is the most natural parallel programming model for vectorized high level languages because it is implemented using arrays, which are often the core data type in a high level language. Briefly, the distributed array model creates distributed arrays in which each process ($P_{ID} \in \{0,...,N_P-1\}$) stores or owns a piece of the whole array.  Additional information is stored in the array so that every $P_{ID}$ knows which parts of the array the other $P_{ID}$s have. How the arrays are broken up among the $P_{ID}$s is specified by a map~\cite{lebak2005parallel, Kepner2009, byun2023ppython}.  For example, Figure~\ref{fig:GlobalArrayMaps} shows a matrix broken up by rows, columns, rows and columns, and columns with some overlap.

  Computations on distributed matrices are usually performed using the ``owner computes'' rule, which means that each $P_{ID}$ is responsible for doing a computation on the data it is storing locally. Maps can become quite complex and express virtually arbitrary distributions.  One example of this complexity is the case in which
a boundary of an array is required by more than one $P_{ID}$ and will be implicitly communicated to complete the computation.  Another example are pipelines which can be implemented by mapping different arrays to different sets of  $P_{ID}$s.

\section{Matrix Multiplication}


The matrix multiplication benchmark constructs several large matrices and multiplies them. Because the amount of computation performed per matrix element is large, performance is primarily constrained by the underlying hardware arithmetic performance.

The matrix multiplication benchmark is defined over three real (or complex) matrices chosen so that their total storage occupies a significant fraction of the available processor memory. The basic algorithm for matrix multiplication is shown in Algorithm~\ref{alg:SerialMatmul} (see caption for details).

\begin{algorithm}[Serial Matrix Multiplicaiton]
The algorithm creates three $N^{2}$ element matrices and performs multiplication on those matrices.
\label{alg:SerialMatmul}
\begin{codebox}
\Procname{$\proc{MatMul}({\bf A}, {\bf B}, {\bf C} : \mathbb{R}^{N{\times}N})$}
\zi           {\bf C} = {\bf A} * {\bf B}
          \End
\end{codebox}
\end{algorithm}

Although matrix multiplication is conceptually simple, it serves as a useful barometer for estimating the performance of many programs on a given hardware platform. The goal of this benchmark is to measure performance in floating point operations per second. It constructs matrices using various data types to exploit the low precision arithmetic units available in recent GPU hardware, thereby demonstrating the significant performance improvements enabled by modern GPU designs.

\begin{figure}[]
\centerline{\includegraphics[width=1.0\columnwidth]{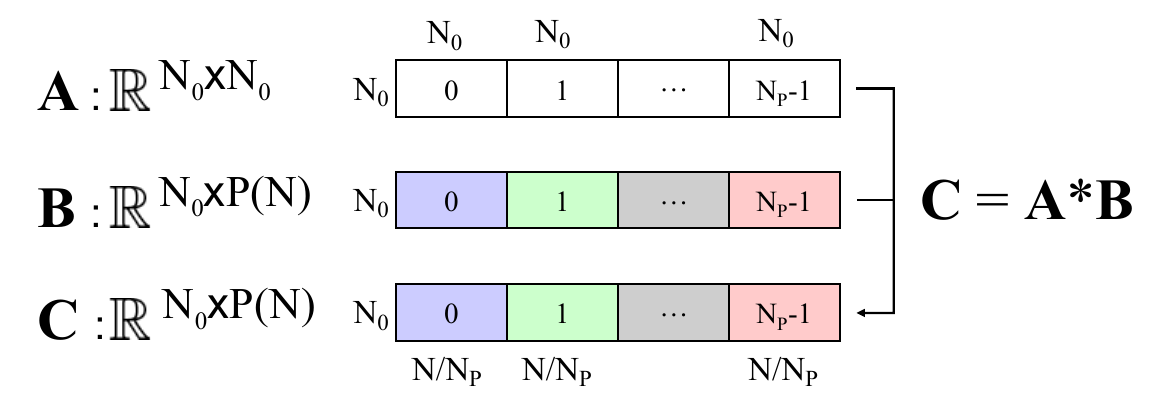}}
\caption{{\bf Parallel Matrix Multiplication Design.}
{\bf A} is a normal $N_0{\times}N_0$ matrix replicated across all processes $P_{ID} \in \{0,...,N_P-1\}$.  {\bf B} and {\bf C} are $N_0{\times}N$ distributed matrices with $N_0{\times}N/N_P$ local matrix on each process.  If {\bf B} and {\bf C} have the same parallel map, then the resulting program will require no communication.}
\label{fig:MatMulDesign}
\end{figure}

The parallel matrix multiplication program can be implemented quite easily using distributed arrays (see Figure~\ref{fig:MatMulDesign}).  As long as all the arrays use the same distributed mapping, the resulting program requires no interprocessor communication, since each of the $N_{p}$  processes is designed to perform an independent matrix multiplication on its own local matrices. The no communication requirement can be strictly enforced by making the program use the $.loc$ construct which specifies that only the part of the array that is local to the $P_{ID}$ is used.  This approach is generally recommended as it guarantees that no communication will take place.  If a mapping that required communication were accidentally used, then the resulting program will either produce an error or will fail to validate. This mapping results in the parallel algorithm shown using $P()$ notation in Algorithm~\ref{alg:ParallelMatMul}, which specifies which dimensions of the array are to be divided among different $P_{ID}$s.   A visual representation of the design for the parallel matrix multiplication algorithm is shown in Figure~\ref{fig:MatMulDesign}.

\begin{algorithm}[Parallel Matrix Multiplication]
The parallel algorithm creates two $N_0{\times}P(N)$ distributed arrays and one $N_0{\times}N_0$ local arrays and performs multiplication operations on just the local part of those arrays. $P()$ denotes the dimension being divided amongst different $P_{ID}$.
\label{alg:ParallelMatMul}
\begin{codebox}
\Procname{$\proc{MatMul}({\bf A} : \mathbb{R}^{N_0{\times}N_0} , {\bf B}, {\bf C} : \mathbb{R}^{N_0{\times}P(N)})$}
\zi           {\bf C}.loc = {\bf A} * {\bf B}.loc
          \End
\end{codebox}
\end{algorithm}

\noindent Typically $N = N_0 N_P$ so the amount of work per process remains constant as the number of processes $N_P$ is increased.

\subsection{Code}
  Code excerpts for distributed array matrix multiplication implementations in Matlab and Python are shown in Code Listings \ref{code:ParallelMatMulMatlab} and \ref{code:ParallelMatMulPython}.  The parallel code can be implemented by simply converting the matrices from regular arrays to distributed arrays.  This same code is run by every instance with the only difference between these instances is that each has a unique $P_{ID}$ that is used to determine which parts of a distributed array belong to each instance (i.e., ${\bf B}.loc$ and ${\bf C}.loc$).

The descriptions of Code Listings \ref{code:ParallelMatMulMatlab} and \ref{code:ParallelMatMulPython} are as follows:

\noindent {\bf Lines 1--3} of the Python code import the required Python packages.  The CuPy package highlighted in blue enables running on GPUs.

\noindent {\bf Line 4} of the Python code defines the data type of the distributed arrays and other variables used in the code. This is a convenient way to varying the precision of the matrices used in the matrix multiplication so that we can experiment the performance with different precision, especially low precision on GPUs.  As written, this Matlab code requires additional functions ({\tt single()}, {\tt half()}, {\tt complex()}) be  applied to each matrix to convert to these types. 

\noindent {\bf Line 9} creates an alternating vector of 1 and -1 for initializing {\tt A} so that resulting matrix multiply stays within the dynamic range of lower precisions.

\noindent {\bf Line 11} creates the parallel map {\tt BCmap} specifying that the $N_{0}{\times}N$ element distributed arrays are to have their columns divided equally among all {\tt Np} over processes with $P_{ID}$ = {\tt 0}, ..., {\tt Np-1}.

\begin{CodeListing}[{\bf Matlab Parallel Matrix Multiplication}]
\label{code:ParallelMatMulMatlab}
\begin{codebox}
\\
\li 
\li 
\li 
\li
\li {\tt\footnotesize lgN0 = 15;  lgN = 15;}
\li {\tt\footnotesize N0 = 2.$^\wedge$lgN0; N = Np .* 2.$^\wedge$lgN; }
\li {\tt\footnotesize A0 = 1.0/N0; B0 =1.0; C0 = 1.0;}
\li {\tt\footnotesize AsignVec = ones(1,N0);}
\li {\tt\footnotesize AsignVec(2:2:end) = -1.0;}
\li 
\li {\tt\footnotesize BCmap = map([1 Np],\{\},0:Np-1);}  \>\>\>\>\>\>\>\>\>\> {\tt\footnotesize \%  Map.}
\li {\tt\footnotesize \% Allocate and initialize matrices.}
\li {\tt\footnotesize     A = (zeros(N0, N0)+A0) .* AsignVec;}
\li {\tt\footnotesize     Bloc = local(zeros(N0, N, BCmap))+B0;}
\li {\tt\footnotesize     Cloc = local(zeros(N0, N, BCmap))+C0;}
\li {\tt\footnotesize     A = {\color{blue} gpuArray}(A);}
\li {\tt\footnotesize     Bloc = {\color{blue} gpuArray}(Bloc);}
\li {\tt\footnotesize     Cloc = {\color{blue} gpuArray}(Cloc);}
\li \> {\tt\footnotesize     tic;}
\li \> \> {\tt\footnotesize         Cloc = A * Bloc;} \>\>\>\>\>\>\>\> {\tt\footnotesize \% Multiplication.}
\li \> {\tt\footnotesize     t = toc;}
\end{codebox}
\end{CodeListing}

\begin{CodeListing}[{\bf Python Parallel Matrix Multiplication}]
\label{code:ParallelMatMulPython}
\begin{codebox}
\\
\li {\tt\footnotesize import numpy as np}
\li {\tt\footnotesize {\color{blue} import cupy as cp}}
\li {\tt\footnotesize import pPython as GPC}
\li {\tt\footnotesize DT=np.float64}
\li {\tt\footnotesize lgN0 = 15;  lgN = 15}
\li {\tt\footnotesize N0 = 2.**lgN0; N = Np .* 2.**lgN}
\li {\tt\footnotesize A0 = 1.0/N0; B0 =1.0; C0 = 1.0}
\li {\tt\footnotesize AsignVec = {\color{blue}cp.ones}((N0))}
\li {\tt\footnotesize AsignVec[1::2] = -1.0}
\li {\tt\footnotesize AsignVec = AsignVec.astype(DT)}
\li {\tt\footnotesize     BCmap = Dmap([1,Np],\{\},range(Np))}  \>\>\>\>\>\>\>\>\>\>\> {\tt\footnotesize \#  Map.}
\li {\tt\footnotesize \# Allocate and initialize distributed matrices.}
\li {\tt\footnotesize A =({\color{blue}cp.zeros}((N0,N0),dtype=DT)+A0)*AsignVec}
\li {\tt\footnotesize Bloc=local(zeros(N0,N,map=BCmap,{\color{blue} gpu=1},dtype=DT)}
\li {\tt\footnotesize Cloc=local(zeros(N0,N,map=BCmap,{\color{blue} gpu=1},dtype=DT)}
\li
\li {\tt\footnotesize Bloc = Bloc + B0}
\li {\tt\footnotesize Cloc = Cloc + C0}
\li \> {\tt\footnotesize     tic = timer()}
\li \> \> {\tt\footnotesize         Cloc = A @ Bloc} \>\>\>\>\>\>\>\> {\tt\footnotesize \# Multiplication.}
\li \> {\tt\footnotesize     t = timer()-tic}
\end{codebox}
\end{CodeListing}

\noindent {\bf Lines 12--18} create a normal array {\tt A} and distributed arrays {\tt B} and {\tt C}; extract their local parts; and then initialize them to {\tt A0}, {\tt B0}, and {\tt C0}.  Because the program requires no communication, the distributed arrays {\tt B} and {\tt C} are never actually allocated.  
Instead, only their local parts {\tt Bloc} and {\tt Cloc} are created. This simplifies the program and reduces the total memory required to run the program by eliminating unnecessary duplication. The Matlab Parallel Computing Toolbox (PCT) function {\tt {\color{blue} gpuArray}()} and the Python CuPy functions {\tt {\color{blue} cp.ones}()}, {\tt \color{blue} cp.zeros}, and {\tt {\color{blue} cp.array}()} (used by pPython {\tt zeros()}), copy  {\tt A}, {\tt Bloc}, and {\tt Cloc} to the GPU memory; subsequent operations on these variables will be automatically performed on the GPU.

\noindent {\bf Lines 19--21} execute the matrix multiplication operations on {\tt Aloc} and {\tt Bloc}, and store the results in {\tt Cloc}.  The time for the matrix  multiplication operation is summed for each trial.  For GPUs a PCT {\tt \color{blue} wait}  command or CuPy {\tt \color{blue} synchronize} command may need to be called prior to each {\tt tic} or {\tt toc} command to ensure the operations are completed on the GPU.

In terms of the differences between Matlab and Python, our preliminary observation is that Matlab makes more choices on behalf of the user to significantly lower the barrier to entry.  Matlab GPU programming is significantly easier than Python GPU programming.  The Matlab performance profiler dramatically simplifies both CPU and GPU performance tuning. There is no equivalent to the Matlab performance profiler in Python.  That said, Python tends to expose more performance tuning knobs which can be used to increase performance.   In general, we used Matlab to find a good implementation and then converted the design to Python.

\section{Application Benchmarks}

\begin{figure}[]
\centerline{\includegraphics[width=1.0\columnwidth]{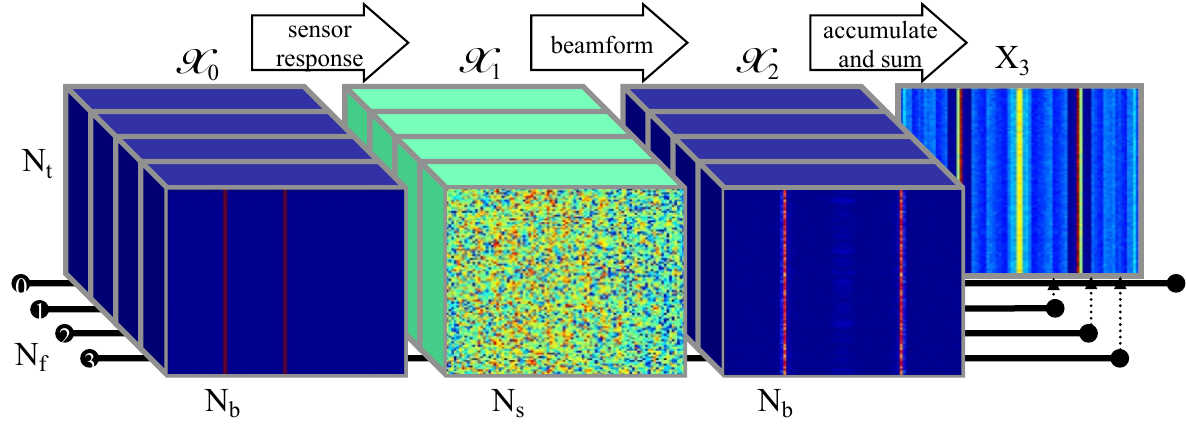}}
\caption{{\bf Parallel Beamformer Design.}
Each stage of the algorithm interacts primarily with two arrays: one for input and one for output. The distributed arrays ${{X}}_{0,1,2}$ are all decomposed
in the $N_{f}$ dimension so that no communication is required in the sensor response, beamform, and write data stages. 
The final stage, accumulation and summation, requires that all results be sent back to $P_{ID} = 0$.  $N_t$, $N_s$, and $N_b$ set the sizes of the various arrays used in the calculation.
}
\label{fig:BeamformerDesign}
\end{figure}

We also examined two application benchmarks, Beamformer and Blurimage.   These benchmarks are representative of real-world sensor applications used across multiple domains. The Beamformer benchmark involves complex matrix multiplications and serves as an essential kernel in many applications. A visual representation of the parallel Beamformer design is shown in Figure~\ref{fig:BeamformerDesign}.

\begin{figure}[h]
\centerline{\includegraphics[width=1.0\columnwidth]{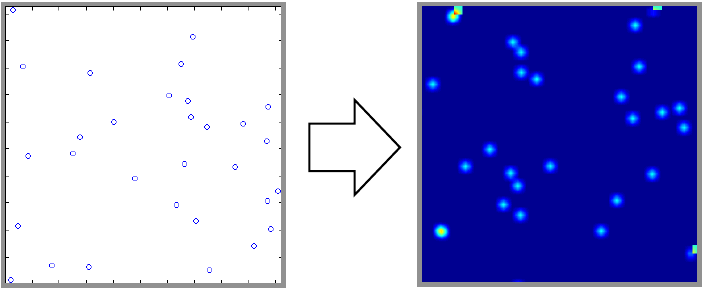}}
\caption{{\bf Blurimage Input and Output.}  The input $N_x{\times}N_y$ matrix consists of a set of random points that then convolve with a $N_{\rm kern}{\times}N_{\rm kern}$ kernel of all 1's. }
\label{fig:BlurimageDesign}
\end{figure}

The Blurimage benchmark is a simple image-filtering program that repeatedly convolves an image with a filter (or kernel) using 2D convolution. This is one of the most common operations in image processing. An example of the input and output images produced by the Blurimage application is shown in Figure~\ref{fig:BlurimageDesign}.   The input  $N_x{\times}N_y$ matrix consists of a set of random points that are then convolved with a $N_{\rm kern}{\times}N_{\rm kern}$ kernel of all 1's. A checkerboard patten of 1 and -1 values can be added to the input so that the resulting matrix stays within the dynamic range of lower precisions.  Convolution can be implemented using both direct and spectral (FFT) algorithms.  Our preliminary experiments indicate that FFT convolution is significantly faster than direct convolution even using optimized direct convolvers from AI libraries.  Thus, all subsequent results reported here are for FFT based convolution.

Operation counts for vector addition, matrix multiplication, and 2D convolution operations are presented in Table~\ref{tab:OperationCounts} for standard and alternative memory intensive algorithms using sparse or spectral (FFT) methods.  Where applicable, the constant factor for multiply accumulate on real values is 2 and for complex values it is 8.  As shown in Table~\ref{tab:OperationCounts} the alternative algorithms provide significant benefits for matrix and 2D convolution operations.  It is worth mentioning that it has become popular to report the performance of accelerator hardware with SpFlops (Sparse Flops), which uses dense operation counts when performing sparse operations.  In some sense this gives credit for work that was not done.  However, it has also become a popular way to highlight algorithmic improvements in a way that is more broadly appreciated.  The same approach can also be applied to 2D convolutions.  Fortunately, it is relatively easy to approximately convert between the two approaches by multiplying/dividing by the appropriate ratio (see column 4 of Table~\ref{tab:OperationCounts}).

\begin{table}
\caption{Operation Counts}
\vspace{-0.25cm}
Standard operation counts for $N$ element vector addition, multiplication of $N{\times}$ matrices, and convolution of a $N{\times}$ matrix with a $N_{\rm kern}{\times}N_{\rm kern}$.  Operation counts are also provided for alternative memory intensive algorithms use sparse or spectral (FFT) methods.  The reduction in operation count is the ratio of the standard to the alternative.
\begin{center}
\includegraphics[width=\columnwidth]{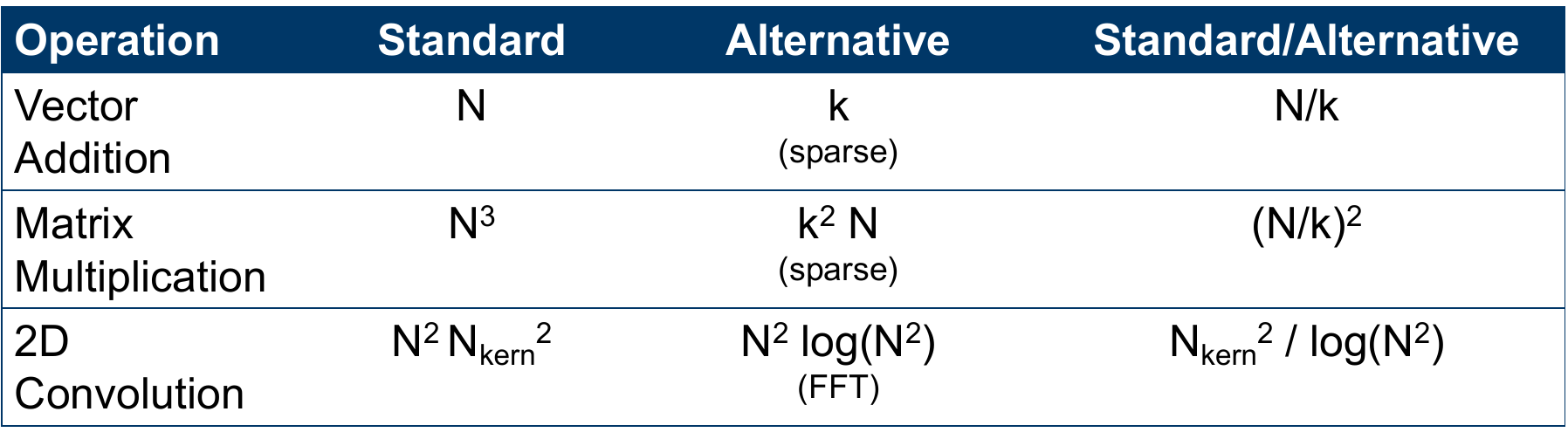}
\end{center}
\label{tab:OperationCounts}
\end{table}%

\section{Computer Hardware}

Our team has developed a high-productivity scalable platform---the MIT SuperCloud---for providing scientists and engineers the tools they need to analyze large-scale dynamic data \cite{kepner2012dynamic, gadepally2018hyperscaling, 8547629}.  The MIT SuperCloud provides interactive analysis capabilities  accessible from high level programming environments (Python, Julia, Matlab) that scale to thousands of processing nodes.  MIT SuperCloud maintains a diverse set of hardware running an identical software stack that allows direct comparison of hardware from different eras.

\begin{table}
\caption{Computer Hardware Specifications}
\vspace{-0.25cm}
MIT SuperCloud maintains a diverse set of hardware running an identical modern software stack providing a unique platform for comparing performance over different eras. GPUs are listed below their host systems. NVIDIA B200 and RTX PRO 6000 GPU systems are hosted at the MGHPCC (Massachusetts Green High Performance Computing Center)  AI Computing Resource (AICR).
\begin{center}
\includegraphics[width=\columnwidth]{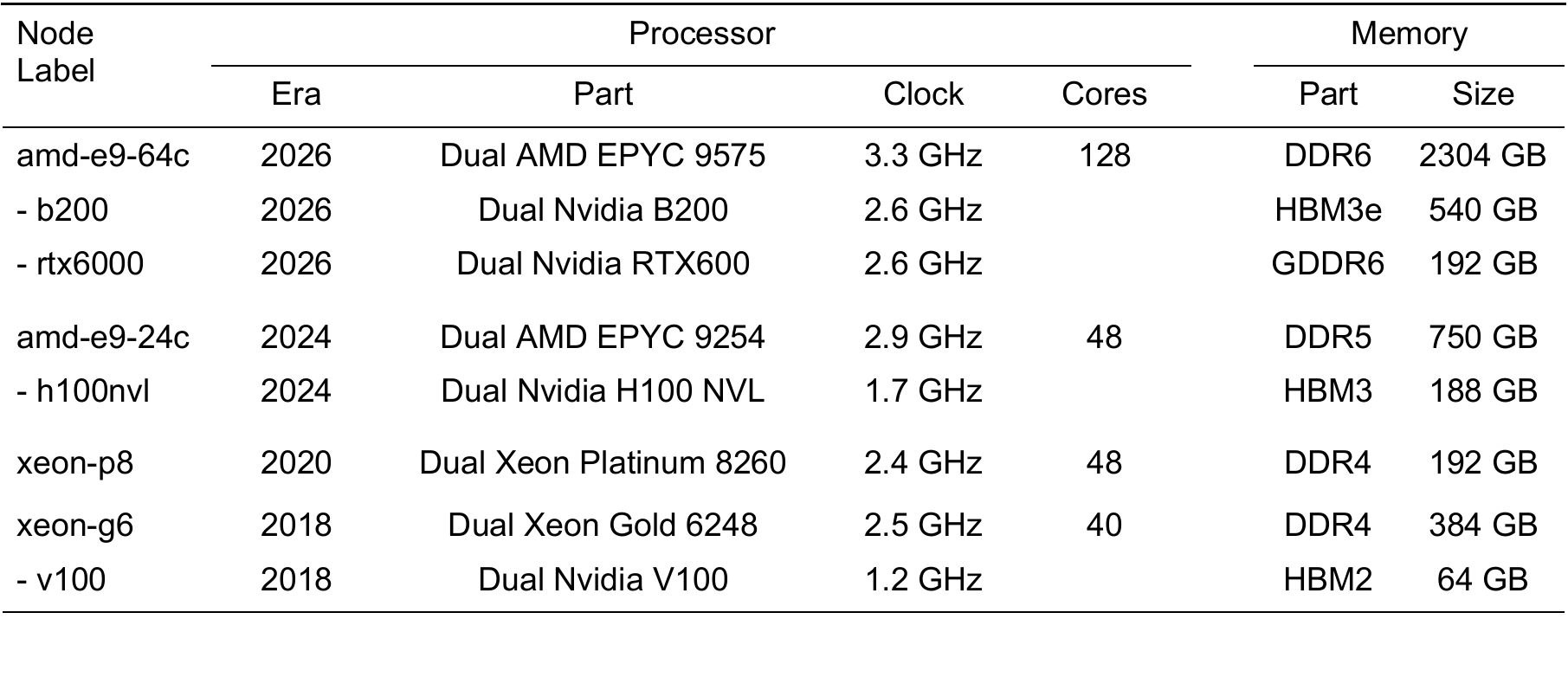}
\end{center}
\label{tab:HardwareTable}
\end{table}%

\begin{table}
\caption{Single Node Benchmark Parameters}
\vspace{-0.25cm}
Benchmark parameters were chosen to balance consistency with the memory capacity of the different hardware. Parameters with $/N_P$ show the dimension that was scaled with increasing number of processes to keep the relative problem size constant for each process. \begin{center}
\includegraphics[width=\columnwidth]{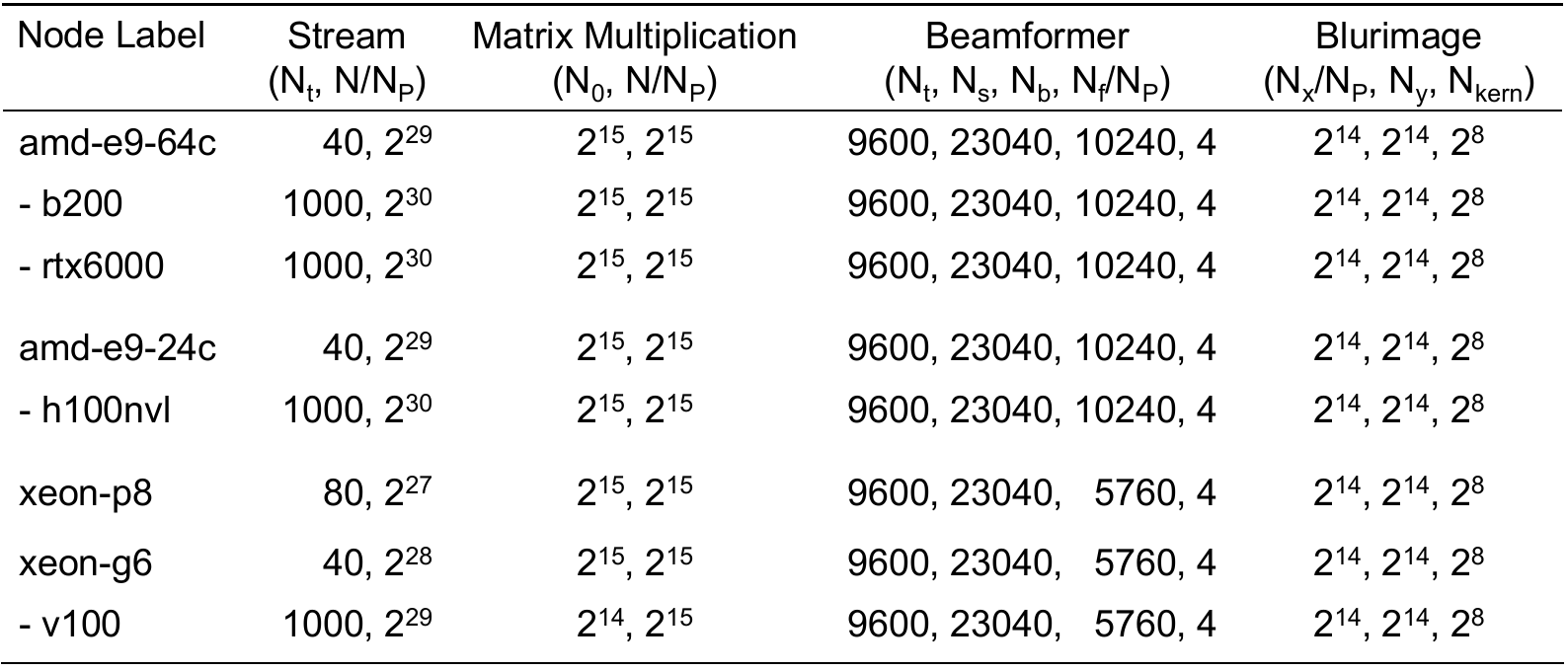}
\end{center}
\label{tab:BenchmarkParameters}
\end{table}%

A typical benchmarking run can be launched in a few seconds using the MIT SuperCloud triples-mode hierarchical launching system \cite{8547629} enabling rapid interactive benchmarking.  The launch parameters were [$N_{node}$ $N_{ppn}$ $N_{tpp}$], corresponding to $N_{node}$ nodes, $N_{ppn}$ Matlab or Python processes per node, and $N_{tpp}$ OpenMP threads per process.   The total number of processes is given by $N_P = N_{node} N_{ppn}$. On each node, each of the $N_{ppn}$ processes and their corresponding $N_{tpp}$ threads were pinned to adjacent cores to minimize interprocess contention and maximize cache locality for the OpenMP threads \cite{byun2019optimizing}.  Within each Matlab or Python process, the OpenMP parallelism is used as provided by their math libraries.  At the end of the processing the results were aggregated using asynchronous file-based messaging \cite{byun2019large}. Triples mode makes it easy to explore vertical scaling by examining combinations of processes and threads on a node, and temporal scaling by running on diverse hardware from different eras.

The computing hardware includes many different types of nodes that have been deployed over nearly a decade (see Table~\ref{tab:HardwareTable}).  All nodes are multicore x86 compatible.  The MIT SuperCloud maintains the same modern software across all nodes, which allows for direct comparison of hardware performance differences.  NVIDIA B200 and RTX PRO 6000 GPU systems are hosted at the MGHPCC (Massachusetts Green High Performance Computing Center)  AI Computing Resource (AICR).

Matlab and Python benchmark distributed array implementations for the different hardware configurations (see Table~\ref{tab:HardwareTable}) were run with the parameters listed in Table~\ref{tab:BenchmarkParameters}.  The size parameters $N$, $N_0$, $N_s$, $N_b$, $N_f$, $N_x$, and $N_y$ were chosen with a base value which is the maximum PCT and CuPy currently allow for GPU calculations.  Within a node $N$ was then scaled with $N_P$ (i.e., constant $N/N_P$) to the maximum memory allowed on the node for Stream and matrix multiplication. A similar approach was used for scaling $N_f/N_P$ and $N_x/N_P$ for Beamformer and Blurimage. Once the node memory maximum was hit, $N$ was kept constant  (i.e., decreasing $N/N_P$) and $N_t$ was increased to keep the relative run time around a few hundred seconds.   

\section{Performance Results}

\begin{table*}[htp]
\caption{Benchmark Results.} Stream results are reported in GBytes/s. All other benchmarks are reported in GFlop/s.  Blurimage (FFT) results use direct convolution operations counts (see Table~\ref{tab:OperationCounts}) to highlight the algorithmic performance gains; underlying FFT operation counts can be estimated by dividing by $N^2_{\rm kern}/(\log(N_x)\log(N_y))$.
The following labeling convention is used:  
\textbf{bold} much better than FP64 MatMul, 
\textbf{\textcolor{blue}{blue}} best precision, 
\textbf{*} TF32 enabled, 
\textbf{**} beyond typical user, 
\textcolor{gray}{$\ll$FP64} much slower than FP64 MatMul, 
\textcolor{gray}{u} unsupported.
\begin{center}
\includegraphics[width=1.5\columnwidth]{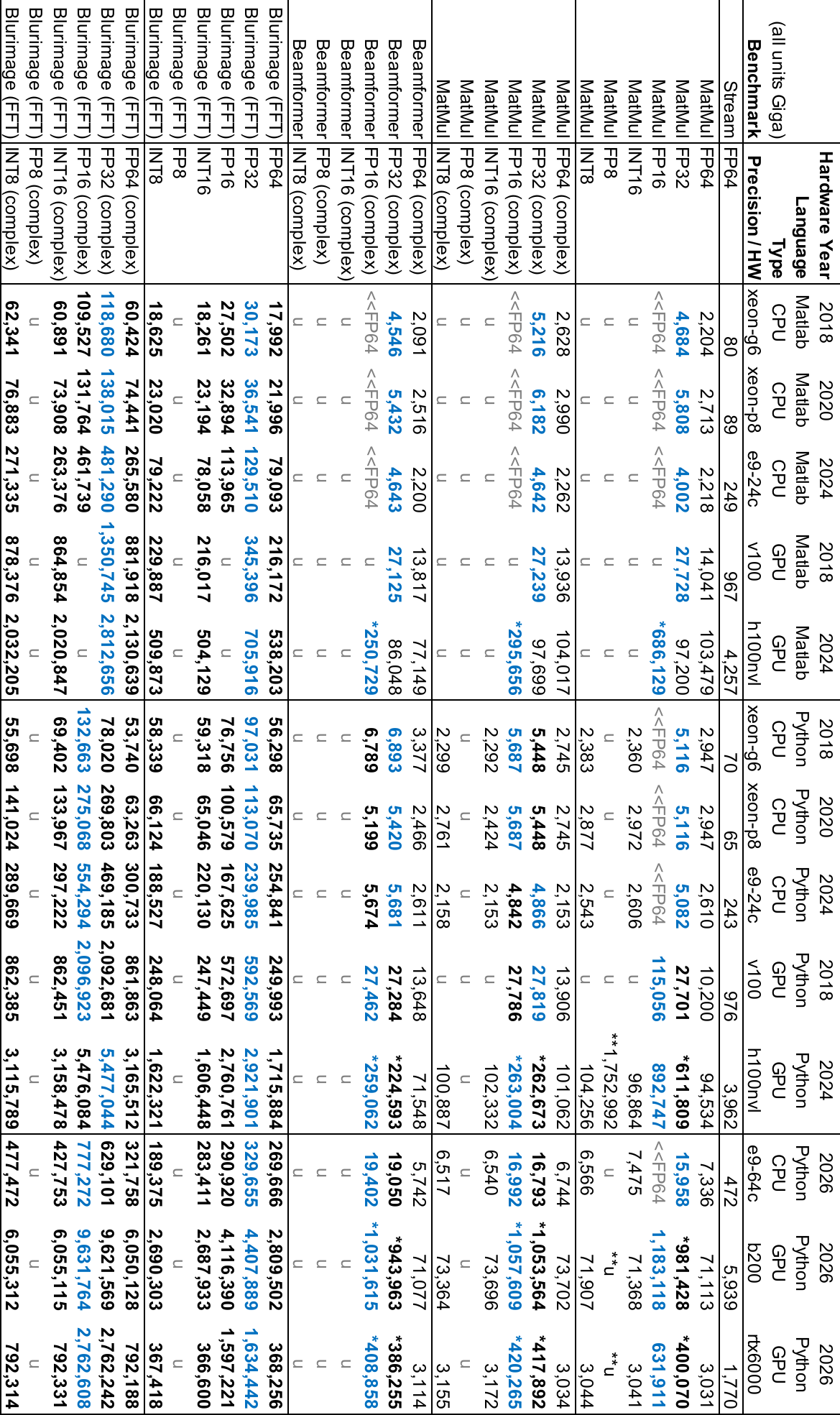}
\end{center}
\label{tab:GPUBenchmarking}
\end{table*}

Table~\ref{tab:GPUBenchmarking} presents the best aggregated within-node benchmark results for all hardware configurations listed in Table~\ref{tab:HardwareTable}.  Stream results are reported in GBytes/s. All other benchmarks are reported in GFlop/s.  Blurimage (FFT) results use direct convolution operations counts (see Table~\ref{tab:OperationCounts}) to highlight the algorithmic performance gains; underlying FFT operation counts can be estimated by dividing by $N^2_{\rm kern}/(\log(N_x)\log(N_y))$.
The following labeling convention is used:  
\textbf{bold} much better than FP64 MatMul, 
\textbf{\textcolor{blue}{blue}} best precision, 
\textbf{*} TF32 enabled, 
\textbf{**} beyond typical user, 
\textcolor{gray}{$\ll$FP64} much slower than FP64 MatMul, 
\textcolor{gray}{u} unsupported.

In general, accessible performance near the underlying hardware limits is available for real/complex FP64/FP32 across all the benchmarks with FP32 offering significant performance boosts over FP64.  When TF32 performance is accessible it offers significant additional performance.  These performance benefits are available with simple vectorized code that runs well on both CPUs and GPUs.  FP16 performance via TF32 is significantly faster, but otherwise generally performs at a FP32 or FP64 level or in some cases is much slower.  INT16/FP8/INT8 is more sporadically supported, when supported it generally provides FP64 level performance.  We did explore what is required to achieve fully FP8 performance on the h100nvl from Python and while significant performance was achieved, the methods used are generally beyond a typical Python user.

More specifically, the Stream benchmark performs well across all systems in both Matlab and Python.  Good Stream performance is often very important in high level languages that tend to do more data copies during the course of execution that low-level languages. Matrix multiplication (MatMul) generally performs as described above.  The main observation is that FP32 is offers significant performance boost and FP16 offers even greater performance when TF32 can be used.  Beamformer tracks complex matrix multiplication and shows that matrix multiplication is accessible within real applications contexts.  Perhaps most interesting is the significant performance achievable with Blurimage (FFT).   This highlights the potential compute intensive vs memory intensive algorithmic tradeoffs now possible with very large and very fast memory.

\section{Conclusion}

 A wide range of applications are currently being enabled by high level programming languages and GPU accelerators.  Retaining productivity requires effective abstractions for achieving performance within a compute node (vertical scaling) and over different generations of hardware (temporal scaling).   One such abstraction that enables high level programming to achieve highly scalable performance are distributed arrays.  By deriving parallelism from data locality, which naturally leads to high memory bandwidth efficiency, distributed arrays provide a straightforward approach for achieving parallel performance.  Using the matrix multiplication and two applications, Beamformer and Blurimage, benchmarks, distributed array performance has been explored on a variety of hardware.   Direct comparison of hardware improvements in combined GPU memory bandwidth and capacity over this period shows a roughly 70× increase over nearly a decade. Over the same interval, the matrix multiplication benchmark improved by about 35×, Beamformer by about 32×, and Blurimage by about 8×. With substantial effort, we achieved matrix multiplication performance exceeding 1 PetaFlop for FP8 on an h100nvl GPU. Future work will seek to extend the distributed array programming model to horizontal scaling across computing nodes and other high level languages and other hardware platforms.
 
\section*{Acknowledgement}

The authors wish to acknowledge the following individuals for their contributions and support:  Guillermo Morales, Bob Bond, Alan Edelman, Peter Fisher, John Goodhue, Jeff Gottschalk, Chris Hill, Joss Knight, Charles Leiserson, Kristen Malvey, Chasen Milner, Sanjeev Mohindra, Sandy Pentland, Heidi Perry, Christian Prothmann, Steve Rejto, Scott Ruppel, Daniela Rus, Pete Saunders, Matt Schuyler, Mark Sherman, Scott Van Broekhoven.




\bibliographystyle{ieeetr}

\bibliography{EasyMultiPrecisionAcceleration}
%

\end{document}